\documentclass[10pt]{article}

\usepackage{amsmath,amssymb,amsfonts,amsthm,textcomp,graphicx,nicefrac,mathtools,mathrsfs, dsfont} 
\usepackage{tikz}
\usepackage{enumitem}
\usepackage{bbm}
\usepackage[utf8]{inputenc}
\usepackage{enumitem}

\def\<{{\langle}}
\def\>{{\rangle}}

\newcommand{\dd}{\mathrm{d}}

\newcommand{\tpose}{\top}
\def\reals{\mathbb{R}} 
\renewcommand{\exp}[1]{\operatorname{exp}\left(#1\right)} 

\providecommand{\diag}{\mathop\mathrm{diag}}

\def\E{\mathbb{E}} 

\def\P{\mathbb{P}} 

\newtheoremstyle{dotless}{}{}{\itshape}{}{\bfseries}{}{ }{}
\theoremstyle{dotless}

\theoremstyle{plain}
\newtheorem{myth}{Theorem}

\newtheorem{mylem}[myth]{Lemma}

\newtheoremstyle{named}{}{}{\itshape}{}{\bfseries}{.}{.5em}{#1 #3}
\theoremstyle{named}
\newtheorem*{namthm*}{Theorem}

\usepackage{chngcntr}
\makeatletter 
\newenvironment{subtheorem}[1]{%
  \counterwithin*{myth}{parentnumber}
  \def\subtheoremcounter{#1}%
  \refstepcounter{#1}%
  \protected@edef\theparentnumber{\csname the#1\endcsname}%
  \setcounter{parentnumber}{\value{#1}}%
  \setcounter{#1}{0}%
  \expandafter\def\csname the#1\endcsname{\theparentnumber\alph{#1}}   
  \ignorespaces
}{%
  \setcounter{\subtheoremcounter}{\value{parentnumber}}%
  \counterwithout*{myth}{parentnumber} 
  \ignorespacesafterend
}
\makeatother
\usepackage[colorlinks=true, citecolor = blue]{hyperref}
\usepackage{cleveref}

\crefname{myth}{Theorem}{Theorems} 

\newcounter{parentnumber}

\usepackage[
backend=biber,
style=alphabetic,
sorting=anyt,
maxbibnames=99,
]{biblatex}
\let\clsCenter\Center\let\clsendCenter\endCenter
\let\Center\undefined\let\endCenter\undefined
\usepackage{ragged2e}
\let\Center\clsCenter
\let\endCenter\clsendCenter

\begin{document}

\title{Lower Bounds for Two-Sample Structural Change Detection in Ising and Gaussian Models}

\author{ {Aditya Gangrade, Bobak Nazer, and Venkatesh Saligrama}\\ {Boston University}\\ 
\{\texttt{gangrade,bobak,srv}\}\texttt{@bu.edu} \footnote{A. Gangrade was supported by DHS Contract: HSHQDC-15-C-B0003 and NSF grant CCF-1618800. B. Nazer was supported by NSF grant CCF-1618800. V. Saligrama was supported by NSF grant CCF-1320547.} }
\date{\vspace{-20pt}}

\maketitle

\begin{abstract}
    The change detection problem is to determine if the Markov network structures of two Markov random fields differ from one another given two sets of samples drawn from the respective underlying distributions. We study the trade-off between the sample sizes and the reliability of change detection, measured as a minimax risk, for the important cases of the Ising models and the Gaussian Markov random fields restricted to the models which have network structures with $p$ nodes and degree at most $d$, and obtain information-theoretic lower bounds for reliable change detection over these models. We show that for the Ising model, $\Omega\left(\frac{d^2}{(\log d)^2}\log p\right)$ samples are required from each dataset to detect even the sparsest possible changes, and that for the Gaussian, $\Omega\left( \gamma^{-2} \log(p)\right)$ samples are required from each dataset to detect change, where $\gamma$ is the smallest ratio of off-diagonal to diagonal terms in the precision matrices of the distributions. These bounds are compared to the corresponding results in structure learning, and closely match them under mild conditions on the model parameters. Thus, our change detection bounds inherit partial tightness from the structure learning schemes in previous literature, demonstrating that in certain parameter regimes, the naive structure learning based approach to change detection is minimax optimal up to constant factors.
\end{abstract}

\section{Introduction}

Markov random fields (MRFs) are a popular way to model the dependence between a set of random variables. Consider a class of MRFs, $\mathcal{M}$, on $p$ random variables, consisting of probability densities parametrised by a vector $\theta \in \mathbb{R}^{\binom{p}{2}}.$ Recall that such MRFs have distributions such that, for any $i,$ $X_i$ is conditionally independent of $X_j$ for $j$ such that $\theta_{ij} = 0$ given the random variables $\{X_k: k \textrm{ s.t. } \theta_{ik} \neq 0\}.$ Thus, with every $\theta,$ one can associate an undirected graph capturing the dependencies, called the Markov network structure of the distribution,  $G(\theta) = ([1:p], E(\theta))$, where $E = \{(i,j) : \theta_{ij} \neq 0 \}.$ The Ising model and the Gaussian MRF are instances of MRFs for binary and real-valued random variables, respectively, and are employed in a variety of applications. \\

An important problem in statistics is the inverse problem of determining the dependencies between a set of random variables given a set of samples drawn from their joint distributions. In the context of MRFs, let $\mathcal{M}$ be a class of MRFs as before, $P\in \mathcal{M}$ be parametrised by $\theta,$ and the random vector $X = (X_i)_{i \in [1\colon p]}$ be distributed according to $P$. The inverse problem can be stated in two closely related ways -  given $n$ i.i.d.~samples $X^n \sim P^{\otimes n}$, one may wish to either learn  $\theta$, which is the \emph{model selection} problem, or to learn $G(\theta)$, which is the \emph{structure learning} problem. These problems are typically studied in the ``\emph{high-dimensional setting,}'' where $n \ll p$, that is, the number of samples is much smaller than the number of variables involved. A large body of work has focussed on constructing structure learning schemes for the Gaussian and Ising models, including algorithms and regularised estimators, with a particular focus on the technical conditions and the number of samples required for consistent learning (see \cite{anandkumar2012high, meinshausen2006, ravikumar2010high} and references therein for the Gaussian MRF, and  \cite{montanari2009graphical, Bresler:2015:ELI:2746539.2746631, ravikumar2010high, anandkumar2012high_ising}, and references therein for the Ising model). A relatively smaller set of papers has studied the hardness of structure learning in terms of simple graph properties, providing necessary and sufficient information-theoretic conditions on the number of samples required for reliable structure learning (see \cite{WanWaiRam} and \cite{SanWai, TandomShanmugamRavikumarDimakis} for the Gaussian and Ising MRFs, respectively).\\  

Recently, the allied problem of change detection in MRFs has received attention. Let $P_1, P_2$ be some $\theta_1, \theta_2$ parametrised members of $\mathcal{M},$ respectively. The basic question is whether, given $n_1$ samples $X^{n_1} \sim P_1^{\otimes n_1}$, and $n_2$ samples {$\widetilde{X}^{n_2} \sim P_2^{\otimes n_2},$} one can estimate $\mathds{1}\{G(\theta_1) = G(\theta_2)\}$ (\emph{structural change detection}) or $\mathds{1}\{\theta_1 = \theta_2\}$ (\emph{general change detection}) well. In the following, we concentrate on structural change detection, and largely suppress the adjective structural, since all results hold for the general problem as well (and also since including it makes for clunky prose). Structural change detection has both received practical interest, for instance in the study of controlled experiments \cite{zhang2012learning}; in genetics \cite{zhao2014direct, xia2015testing}; and in neuroscientific contexts \cite{belilovsky2016testing}, and theoretical interest focused on change detection algorithms along with the study of their sample complexities and consistency conditions \cite{zhao2014direct, FazBan, liu2017, liu2017learning}. {A related line of work is pursued in \cite{daskalakis2016testing}, which studies, amongst other things, the sample complexity of goodness of fit testing for Ising models under statistical measures of difference, i.e., given $P$ and $X^n \sim Q^{\otimes n},$ where $P,Q$ are Ising models, it studies the problem of estimating $\mathds{1}\{D(P\|Q) + D(Q\|P) \ge \varepsilon\},$ where $D$ is the Kullback-Leibler (KL) divergence. We note that our proof technique also relies on analysing a goodness of fit test, but against structural distance measures. }\\

In this paper, we take an information-theoretic approach to the change detection problem for the Ising and the Gaussian MRFs. Following the structure learning literature, we study the problem for the simple case of distributions that have Markov network structures with maximum degree bounded by $d$, denoted $\mathcal{G}_{p,d}$, and study the same against a minimax risk of misdeclaring a change, or the lack thereof, which is denoted $R_{\textrm{cd}}.$ We allow Ising models where the non-zero $\theta_{ij}$ are lower bounded by some $\alpha$ and upper bounded by some $\beta$, and Gaussian MRFs where the ratios of the non-zero off-diagonal entries to the corresponding diagonal entries are lower bounded by some $\gamma$.  Our results provide lower bounds on the sample complexity in terms of $p,d,$ and $(\alpha, \beta)$ or $\gamma$, for any detection method that makes $R_{\textrm{cd}}$ small, and use simple ensembles of possible changes to do so. Our proof technique is symmetric in the two datasets, and hence the results are lower bounds on $\min(n_1,n_2)$. Interestingly, under mild conditions on the parameters, these bounds improve upon all the known lower bounds on the sample complexity of structure learning, and are at most $\mathrm{poly}(d)$ separated from the sample complexity of the maximum likelihood structure learner. This suggests\footnote{but does not prove} that, at least for the Ising and Gaussian MRFs on $\mathcal{G}_{p,d},$ change detection is as hard, in terms of its data cost, as structure learning.\\

\subsection{Comparison to Prior Work}

Note that one can naively perform change detection by estimating the network structures of the distributions underlying the two sets of samples and comparing the same. This method is generally considered profligate, especially in the pratically important case where the underlying models might be dense, but the differences between them are small\footnote{All the applied work we have mentioned before falls into this category.}. One possible justification for this comes from the compressed sensing literature, which has demonstrated that in certain model classes learning sparse changes between two models can be significantly more data efficient than learning either of the models. Thus, a widely believed ``folk theorem'' asserts the existence of schemes for change detection in the Ising and Gaussian MRFs that can handle wide ranges of model classes such that their sample requirements scale only with sparsity of the changes to be detected, rather than the complexity of the underlying models. The previous work mentioned, \cite{zhao2014direct, FazBan, liu2017, liu2017learning}, all develop algorithms that estimate certain functionals of the ratio of the probability densities of the underlying models via a regularised optimisation scheme, and detect change on the basis of this estimate with sample complexities that scale as $O(s^k\log p)$, where $s := \|\theta_1- \theta_2\|_0$. However, these results only hold under strong technical conditions, {e.g.} structural assumptions on the Fisher information matrices of the models considered, and as previously noted in \cite{montanari2009graphical, anandkumar2012high}, such conditions do not easily provide a clear description of which classes of graphs and models satisfy them in direct terms, that is, do not provide a non-trivial class $\Theta$ such that these results hold for every $\theta \in \Theta$. {Further, as \cite{Hein2016} points out for the case of Gaussian structure learning, such conditions can fail to hold in subtle ways for application relevant data.} \\

The nature of our results is at odds with the conventional wisdom (e.g., the claims in \cite{FazBan, liu2017learning}) that sparse changes can be easy to detect irrespective of the ranges underlying parameters $\theta_1,\theta_2$ and the graph classes they live in. We believe that this disconnect is because of the strictness of the conditions required for these results to hold, and the regimes they implicitly push the underlying models into. In particular, as previously noted, our lower bounds on sample complexity closely match all known lower bounds on structure learning in $\mathcal{G}_{p,d}$ under mild conditions on the parameters. Crucially, the ensembles we construct to demonstrate these bounds have the sparsest possible changes - $\|\theta_1 - \theta_2\|_0 = 1.$ We note that this dependence on the parameters and $d$ is nontrivial - for instance, for the Ising model, if $\beta >1$, then the sample complexity lower bound scales as $\Omega\left( e^{2\beta d} d^{-1} \log p\right)$, which is exponential in $d$, and even if one is allowed to cherry pick $(\alpha, \beta)$ to minimise the lower bound, one needs at least $\Omega\left( \frac{d^2}{(\log d)^2} \log p \right)$ samples to detect change, irrespective of any incoherence or dependency conditions that may be imposed. For the Gaussian, our bound matches the scaling of the corresponding the structure learning bound. Since this is known to be tight up to constant factors in certain regimes, this shows that the naive scheme for change detection is minimax optimal in at least some contexts.\\

Lastly, our lower bounds on Ising model change detection sample complexity for $\mathcal{G}_{p,d}$ structured models are actually stronger than those in \cite{SanWai} in certain parameter regimes. Since change detection is reducible to structure learning, these are also new bounds for the latter. We further note that our proof technique, which is currently un-optimised, relies upon lower bounding the {$\chi^2$-distance of certain distributions,} and thus differs from the Fano bounds of the previous literature. Since the technique is largely elementary and provides improvements on the previous results, it may be of independent interest and merit refinement.\\

\noindent \emph{Organisation:} \S\ref{sec:def} defines the models considered, and precisely formulates the problem considered, while \S \ref{sec:results} states the results and compares them with previous work. \S \ref{sec:proof} begins by laying out the common technical development involved in the proof, and follows this with the actual proofs.\\

\noindent\emph{Notation:} For a natural $n$, we use $[1:n]$ as shorthand for the set $\{1, 2, \dots, n\}$. $X,\widetilde{X}$ denote random vectors, usually of dimension $d$ or $p$, with $\mathcal{X}$-valued entries. For a natural $i,$ $X_i$ is the $i^{\mathrm{th}}$ component of $X$. Similarly, for a set of naturals, $S,$ $X_S$ is the vector $(X_i)_{\{i \in S\}}$.  We use $X^n$ to denote an $n$-length sequence of i.i.d.~random vectors, frequently referred to as a `dataset'. For a distribution $P$, and given $X \sim P$, $P^{\otimes n}$ is the distribution of $X^n$. Further, given $X^n,$ the $t^{\mathrm{th}}$ sample in that dataset is denoted $X^{(t)}$. Lastly, we use $Z$ and diacritical modifications of the same for partition functions whenever they are relevant. The two element sets $\{i,j\}$ are interchangeably denoted $(i,j)$ when referring to edges in an undirected graph, and as just $ij$ when they appear in a subscript. Vectors in $\mathbb{R}^{\binom{n}{2}}$ are indexed by cardinality-two subsets of $[1:n]$. For instance, a vector $\theta \in \mathbb{R}^{\binom{3}{2}}$ is represented as $(\theta_{12}, \theta_{13}, \theta_{23})$. The identity matrix of size $p$ is denoted as $I_p$.\\

For functions $f,g,$ $f = O(g)$ if there exists a positive constant such that $\lim_{n \to \infty} f(n)/g(n) \le C,$ and $f = o(g)$ if $\lim f/g \to 0$. Similarly, $f = \Omega(g)$ if $g = O(f),$ and $f = \omega(g)$ if $g = o(f)$.

\section{Definitions and Problem Statement} \label{sec:def}

We begin the paper with a bit of background on Markov random fields (MRFs) and a precise problem statement, which also serve to establish notation followed in the rest of the paper. Note that while we define the problem over a general MRF, our results are restricted to the Ising and the Gaussian MRFs. We define the problem thus since we develop a single technique for establishing lower bounds on the sample complexity and apply it parallely to both the models considered.

\subsection{Markov Random Fields}
Recall that an undirected finite simple graph $G= (V,E)$ consists of a finite vertex set $V$, here identified with $[1:p]$, and edge set $E$ which is a set of subsets of $V$ of cardinality $2$. For $u \in V$, we let $\partial u := \{v: \{u,v\} \in E\}$ be the set of neighbours of $u$, and the cardinality of $\partial u$ is said to be the degree of $u$. Lastly, for naturals $p,d \le p-1,$ we let $\mathcal{G}_{p,d}$ be the set of graphs on $p$ vertices such that every node's degree is no bigger than $d$.\\

For a set $\mathcal{X},$ an $\mathcal{X}$-valued \emph{Markov random field} on a graph $G$ is a random vector $X = (X_v)_{v \in V} \in \mathcal{X}^{|V|}$ such that for every $v$, $X_v$ is conditionally independent of $X_{V \setminus (\{v\} \cup\partial v)}$ given $X_{\partial v}$. The graph $G$ is said to be the Markov network structure of the Markov random field. We define a class of MRFs as a set of probability distributions, each of which is an MRF. For a class of MRFs $\mathcal{M},$ and a family of graphs $\mathcal{G}$, $\mathcal{M}$ is said to be a class of MRFs on $\mathcal{G}$ if every MRF in $\mathcal{M}$ has a Markov network structure contained in $\mathcal{G}$.\\

Given a graph $G$ and a vector $\theta \in \reals^{\binom{|V|}{2}}$ such that $\{i,j\} \not\in E \Rightarrow \theta_{\{i,j\}} = 0$, a $0$-external field \emph{Ising model} on $G$ with parameter $\theta$ is a $\{\pm 1\}$-valued Markov random field with the distribution \begin{equation}\mathbb{P}_{(G,\theta)}\left(X = x\right) = \frac{1}{Z(\theta)} \exp{ \sum_{\{u,v\}\in E} \theta_{uv} x_ux_v}, \notag\end{equation} where $Z(\theta)$ is a normalising constant commonly known as the partition function. We let $\mathcal{I}_{p,d}(\alpha, \beta)$ be the set of Ising models on graphs in $\mathcal{G}_{p,d}$ such that for every $u,v,$ either $\theta_{uv} = 0$, or $\alpha \le |\theta_{uv}| \le \beta$ holds. Note that every $\theta$ determines a network structure, which we refer to as $G(\theta)$.\\

Similarly, given a graph $G,$ a symmetric, positive-definite matrix $A$ such that $A_{i,j} \neq 0$ only if $(i,j) \in E$, the $0$-mean \emph{Gaussian Markov Random Field} on $G$ with parameter $A$ is the $\reals$-valued Markov random field with the distribution \begin{equation} \P\left( X \in S\right) = \sqrt{\frac{\det A}{(2\pi)^{|V|}}} \int_{x \in S} \exp{-\frac{x^{\tpose} A x }{2}} \,\dd x. \notag\end{equation} The matrix $A$ is known as the precision matrix of $X$. Note that the Markov network structure of the distribution is determined by the non-zero entries of $A$. For $\gamma >0,$ let $\mathcal{N}_{p,d}(\gamma)$ be the set of Gaussian MRFs on graphs in $\mathcal{G}_{p,d}$ with $0$ mean and precision matrix $A$ such that all diagonal entries are non-zero, and for every $(i,j) \in V^2,$ \begin{equation}  \min_{A_{ij} \neq 0} \left|\frac{A_{ij}^2}{A_{ii}A_{jj}}\right| \ge \gamma^2 \notag\end{equation}

We note here that the value of $\gamma$ actually modulates the graphical structures allowed within $\mathcal{G}_{p,d}$. For instance, if we allow graphs such that even a single node may have $d$ neighbours in the Markov network structure of a distribution as above, then the condition of positivity of the precision matrices enforces $\gamma < 1/\sqrt{d}$, and if we allow the entirety of $\mathcal{G}_{p,d}$ unrestrictedly, the condition $\gamma < 1/d$ is required.\\

Lastly, for the above classes of MRFs, we frequently describe distributions in terms of their Markov network structures. In particular, we say that a distribution has the edge $(i,j)$ with weight $w$ if $\theta_{ij} = w$ for the Ising model, and if $A_{ij} = A_{ji} = w$ for the Gaussian MRF. If all the edges of a distribution have the same weight, we say that the distribution has uniform edge weights.

\subsection{The Change Detection Problem}

Let $\mathcal{M}$ be a class of $\mathcal{X}$-valued Markov random fields with parameters $\theta \in \Theta$, and let $P_1, P_2 \in \mathcal{M},$ have parameters $\theta_1$ and $\theta_2$ respectively. Note that the distributions/parameters are unknown to us, and are potentially equal. For $P\in\mathcal{M}$, let $G(P)$ be its Markov network structure. We consider the structural change detection problem, which may be informally stated as follows: Given $n_1$ samples drawn according to $P_1$ independently and identically, and $n_2$ samples identically drawn according to $P_2$ independently of each other and of the other set of samples, can one determine if $G(P_1) = G(P_2)$ or not with high probability?\\

Formally, for $\mathcal{X}$-valued Markov random fields $P_1, P_2 \in \mathcal{M},$ let $X^{n_1} \sim P_1^{\otimes n_1},$ and $\widetilde{X}^{n_2} \sim P_2^{\otimes n_2}$ be finite sets of samples, also referred to as datasets, drawn independently and identically from $P_1$ and $P_2$, respectively. An $(n_1,n_2)$-change detector for $\mathcal{M}$ is a map $\phi: \mathcal{X}^{n_1} \times \mathcal{X}^{n_2} \to \{0,1\}.$ Let $\mathbf{\Phi}_{n_1,n_2}$ be the set of all $(n_1,n_2)$-change detectors. Let the risk of a detector $\phi$, be \begin{align*}R(\phi;n_1,n_2,\mathcal{M}) := \sup_{ P_1, P_2 \in \mathcal{M}}  \,\,\,\,&\P\left\{\phi(X_1^{n_1}, X_2^{n_2}) = 1 \mid G(P_1) = G(P_2)\right\} \\ +\, &\P\left\{\phi(X_1^{n_1}, X_2^{n_2}) = 0  \mid G(P_1) \neq G(P_2)\right\}.\end{align*}  
and the minimax change detection risk with $(n_1,n_2)$ samples over the class $\mathcal{M}$ be \begin{equation}R_{\mathrm{cd}}(n_1,n_2;\mathcal{M}) := \inf_{\phi \in \mathbf{\Phi}_{n_1,n_2}} R(\phi;n_1,n_2,\mathcal{M}). \notag\end{equation} 

Note that the above risk expressions are just the adaptation of the standard binary hypothesis testing risks to our situation.\\

We say that an $(n_1,n_2)$-change detector is $\delta$-reliable over the class $\mathcal{M}$ if $R_{\textrm{cd}}\left(\phi ; n_1,n_2,\mathcal{M}\right) < \delta,$ and say that the change detection problem can be solved over $\mathcal{M}$ $\delta$-reliably with $(n_1,n_2)$ samples if there exists an $(n_1,n_2)$-change detector over $\mathcal{M}$ that is $\delta$-reliable or, equivalently, if $R_{\textrm{cd}}\left(n_1,n_2;\mathcal{M}\right) < \delta.$ The parameter $\delta$ is occasionally referred to as the reliability level.\\

The main aim of this paper is to study the trade-off between the reliability level $\delta$ of change detection over a given class of MRFs and the sample size $(n_1,n_2).$ In particular, we provide necessary conditions on $(n_1,n_2)$ for $\delta$-reliable change detection with $(n_1,n_2)$ samples over the classes $\mathcal{I}_{p,d}(\alpha,\beta)$ and $\mathcal{N}_{p,d}(\gamma)$ in terms of their parameters.

\section{Theorem Statements and Nature of Results}\label{sec:results}

As previously noted, our results are necessary conditions on the number of samples required for $R_{\textrm{cd}}$ to be smaller than a given $\delta$. The results are stated in separate subsections for the Ising and the Gaussian MRFs, respectively. A discussion comparing the results to parallel results in structure learning follows the theorem statements, while proofs are relegated to later sections. C.f. \S \ref{sec:def} for precise definitions of the models and graph class considered. 

\subsection{Ising Model}

\begin{subtheorem}{myth}\label{thm:ising}
        
    \begin{myth}\label{thm:ising_easy}
            Let $\alpha >0.$ For every $\delta \in [0,1],$ a necessary condition for $\delta$-reliable change detection with $(n_1,n_2)$ samples over $\mathcal{I}_{p,d}(\alpha,\beta)$ is \begin{equation*} \min(n_1,n_2) > \frac{\log\left(1 + 4(1-\delta)^2 \binom{p}{2}\right)}{\log (1 + \tanh^2 \alpha)} .\end{equation*}
    \end{myth}
    \begin{myth}\label{thm:ising_1}
             Let $d \ge 4$ and $ \beta (d-3)\ge \ln d.$ For every $\delta \in [0,1],$ a necessary condition for $\delta$-reliable change detection with $(n_1,n_2)$ samples over $\mathcal{I}_{p,d}(\alpha,\beta)$ is \begin{equation*}
                \min(n_1, n_2) > e^{2\beta d}\frac{\ln\left(1 + 4(1-\delta)^2 \left\lfloor p/(d+1)\right\rfloor\right)}{8(e^{4\beta} + d)}. 
            \end{equation*}
    \end{myth}    
    
\end{subtheorem}

\noindent \emph{Proof sketch:} While the exact proof is relegated to \S\ref{sec:proof}, we loosely detail the strategy here. Let $P \in \mathcal{M}$, and let $\mathcal{Q} \subset \mathcal{M}$ be a set of distributions such that $G(P) \not\in G(\mathcal{Q}).$ Suppose we are given the information that the larger set of samples is drawn according to $P$, while the second set is either drawn from some $Q \in \mathcal{Q},$ or drawn from $P$. We have thus reduced the change detection problem to running a hypothesis test on the smaller dataset, with the simple null hypothesis that the same is drawn from $P$, and the composite alternate that it is drawn according to some $Q \in \mathcal{Q}$. Suppose further that we are supplied with a prior, $\pi_Q$ for the selection on $\mathcal{Q}$. Clearly, the average risk of the hypothesis testing problem under the prior $\pi_Q$ would be a lower bound for the minimax change detection risk. However, since the uniformly most powerful tests for such problems are known by the classical results of Neyman \& Pearson, we can compute lower bounds on these average risks. In particular, we do this by applying a variation of Le Cam's method, as outlined in \cite{arias-castro2012}, and with a uniform prior on $\mathcal{Q}.$\\
            
We call the pair $(P,\mathcal{Q})$ a change detection ensemble. The ensembles used to derive the above bounds are as follows.\begin{itemize}[labelindent=-1pt]  
    \item Theorem~\ref{thm:ising_easy}: $P$ is the Ising model with no edges on $p$ nodes, and $\mathcal{Q}$ is the collection of Ising models with exactly one edge with edge weight $\alpha.$
    \item Theorem~\ref{thm:ising_1}:  $P$ is the Ising model with uniform edge weight $\beta$ on $\lfloor p/d+1 \rfloor$ separate cliques of size $d+1$ each, while $\mathcal{Q}$ is the collection of Ising models on the same graph as $P$ but with one known edge from exactly one of the cliques deleted,  again with uniform weight $\beta.$ 
\end{itemize}

 We note that the above proof technique directly allows us to state our results as structure learning bounds in the context of the recovery criterion defined in \cite{SanWai} as well, as will be discussed in \S \ref{sec:proof}.\\

\emph{Remark:} The ensemble used for the proof of Theorem \ref{thm:ising_1} most likely does not satisfy the conditions required in work such as \cite{FazBan} or \cite{liu2017}. This is because these papers all require an `incoherence condition', and, as pointed out in \cite{montanari2009graphical}, these conditions essentially hold only if $\beta d < k$ for some $k<1$, while we need $\beta d = \Omega(\log d)$ for the results to follow.

\subsubsection{Remarks, and comparison with structure learning bounds}

In the high-dimensional setting, one considers the behaviour of these bounds for large $p$. The above bounds above can be interpreted in three different contexts depending on which of the model parameters are allowed to change. In the following we set $\delta = 1/2,$ in order to discuss conditions necessary to beat a random coin, and $c$ is some arbitrary quantity that depends only on the non-changing parameters. 

\begin{enumerate}
    \item If the parameters $d,\alpha, \beta$ are given constants, then change detection requires at least $c\log p$ samples from each dataset to detect changes.
    \item If we hold $\alpha, \beta$ as constants, and allow $d$ to grow with $p$, then for sufficiently large $d$, one needs $ce^{2\beta d}\frac{\log p/d}{d}$ samples from each dataset to detect changes.
    \item If we allow $\beta$ and $\alpha$ as well as $d$ to change with $p$, then unless $\beta$ decays with $d,$ we are forced into the exponential growth in $d$ regime. Suppose we impose the requirement that the bounds grow at most polynomially in $d$. This can be done essentially by limiting $\beta \le k \log d/d$ for any constant $k$, which limits the second bound, although since $\alpha \le \beta$ must hold, this affects the first bound as well. Optimising for the $k$ which gives the lowest net growth, we get that in \emph{any scenario} of parameter growth, we need at least $ \frac{(d-1)^2}{4 \log^2 d} \log p$ samples from each data set to detect changes.
\end{enumerate}

We compare the bounds of Theorem \ref{thm:ising} with two results due to Santhanam \& Wainwright. Note that the statements have been modified to fit our notation.\\

First we consider the necessary condition: 

\begin{namthm*}[{\cite[Thm. 1]{SanWai}}]
Consider $\mathcal{I}_{p.d}(\alpha, \beta)$ for $\beta d > 1.$ If the structure learning probability of error is smaller than $\delta - \frac{1}{\log {p}},$ then following condition on the number of samples, $n$, is necessary\begin{equation*} n>  (1-\delta)\max \left\{\frac{\log p}{2 \alpha \tanh\alpha}, \frac{e^{\beta d} \log (pd/4 - 1)}{4\beta d e^\beta}, \frac{d}{8} \log \frac{p}{8d}  \right\}. \end{equation*}
\end{namthm*}


Theorem \ref{thm:ising_easy} is the direct analogue of the first bound, and Theorem \ref{thm:ising_1} is the direct analogue of the second bound. The third bound is only active when all parameters are held constant, and even then is inactive for $d > 90$, and thus left largely unconsidered by us. However, the other two terms are weaker than Theorem \ref{thm:ising} in the regime in which the latter hold.  In particular, if $d = o(p),$ and $\beta d> \log d$, Theorem \ref{thm:ising_1} has an advantage of essentially $c e^{\beta d} \frac{\beta d e^{\beta}}{e^{4\beta} + d},$ which is unbounded in $d$ for $\beta > 2 \log d /d.$ {Note that while the $\beta d > \log d$ requirement for our second bound to hold may seem more stringent than the $\beta d > 1$ condition in \cite{SanWai}, in the regime $\beta d < \log d$, the $\alpha^{-2} \log p$ bound dominates the exponential bound in both cases, and thus this distinction is rendered moot.} Lastly, observe that if we force $\alpha$ and $\beta$ to decay in a manner that allows at most polynomial growth with $d$, we see that structure learning requires $\Omega\left(\frac{d^2}{(\log d)^2} \log p\right)$ samples as in our case.\\

In the same paper, Santhanam \& Wainwright also show  (\cite[Theorem 3a)]{SanWai}) that if the edge weights are given, it is possible to learn the structure with $n$ samples for  \begin{equation} n \gtrsim \frac{3(3e^{2\beta d} + 1)}{\sinh^2 (\alpha/2)} d\left(3 \log p  + \log 2d + \log \frac{1}{\delta}\right). \notag \end{equation}  Ignoring the $\delta$ terms, the result above is separated from our lower bound by a factor of $c\frac{d^2 + e^{4\beta} d}{\sinh^2{\alpha/2}}.$ {Now if $1/\alpha$ and $e^{4\beta}$ are at most polynomial in $d,$ then this factor is at most polynomial in $d$, which is neglegible compared to the exponential in $d$ scaling forced under $\beta d = \omega( \log d)$. Since our bounds are computed with a change ensemble where we are aware of the edge weights, the closeness of these bounds implies that our technique cannot yield significantly stronger lower bounds in this regime. Lastly, recall that our change detection bounds can also be stated as structure learning bounds. Thus, our results close the exponential gap in the structure learning lower bounds of \cite{SanWai}, which has persisted through all subsequent work on structure learning.}



\subsection{Gaussian MRFs}
    \begin{myth}\label{thm:Gaussian}
        Let $\gamma \le 0.39$. For every $\delta \in [0,1],$ a necessary condition for $\delta$-reliable change detection with $(n_1,n_2)$ samples over $\mathcal{N}_{p,d}(\gamma)$ is \begin{equation}\min(n_1,n_2) > \frac{1}{2 \gamma^2} \log \left(1 + (1-\delta)^2p\right). \notag\end{equation}
    \end{myth}
    
\emph{Remark:} As mentioned before, $\gamma$ controls the richness of the Markov network structures within $\mathcal{G}_{p,d}$ that are allowed, essentially since the two together determine the positivity of certain precision matrices in the class. In particular, if we allow even a single node to have $d$ neighbours, as we rightly should when considering models in $\mathcal{G}_{p,d}$, then $\gamma < 1/\sqrt{d}$ is forced, and if $d$ regular graphs are allowed, then $\gamma < 1/d$ is imposed. In light of this, the condition on $\gamma$ in the theorem statement is fairly benign. For instance, enforcing $\gamma < 1/\sqrt{d}$ for $d \ge 7$ already gives us $\gamma < 0.38.$\\

As in the Ising case, the bound is proved by considering an explicitly stated restricted class of changes, and bounding the risk for them. Curiously, while we obtain the above bound by considering a simple ensemble of changes of the form independent versus one-edge, essentially the same result can be obtained up to constant factors\footnote{ and subject to the $\gamma$ conditions that allow these ensembles to exists,} in more richly connected classes of ensembles - for instance, by using repetitions as in \S\ref{sec:K_d_extend} of the star graph versus the star graph with one edge moved, or the complete graph versus the same with a known edge missing. This suggests that there might be some uniformity to the hardness of structure learning/change detection in Gaussian MRFs.\\

Comparing the above bound to the corresponding structure learning bound, we note the following result of Wang, Wainwright, and Ramchandran, stated in our notation.
\begin{namthm*}[{\cite[Thm. 1]{WanWaiRam}}] Consider the class $\mathcal{N}_{p,d}(\gamma)$ with $\gamma \in [0,1/2]$. A necessary condition for asymptotically\footnote{as $p$ grows large} reliable structure learning over this class is \begin{equation*} n > \max \left\{ \frac{\log\binom{p-d}{2} - 1}{4\gamma^2} , \frac{2\log \binom{p}{d} - 1}{\log \left(1 + \frac{d\gamma}{1- \gamma} \right) - \frac{d\gamma}{1 + (d-1) \gamma}}\right\} .\end{equation*}\end{namthm*}

Note that the first term in the expression above dominates the second when $\gamma = o(1/\sqrt{d}),$ which, by the previous argument, is the range of $\gamma$ in which at least one node can be connected to $d$ other nodes in the Markov network structure of the graph. Thus, our lower bound matches the structure learning lower bound in the parameter region relevant for $\mathcal{G}_{p,d}$. {We note that this bound is near tight - for instance \cite{anandkumar2012high} achieves, under technical conditions, structure learning for Gaussian MRFs with sample complexity $O(\gamma^{-2}\log p/\delta)$.}

\section{Proofs}\label{sec:proof}

We begin by detailing the general proof technique that we use, followed by the proofs of Theorems \ref{thm:ising_easy} and \ref{thm:Gaussian}, which are of the same flavour, and are rather simple. The proof of Theorem \ref{thm:ising_1} is relatively more involved and follows these sections.

\subsection{Lower Bounding Technique}\label{sec:low_b_tech}

    We first note that any lower bound on sample complexities to achieve a given risk level must hold for both $n_1$ and $n_2$, since merely switching the labels of the two sets of samples should not affect anything. We hence set $n = \min(n_1, n_2)$, and derive lower bounds on $n$ by considering simpler hypothesis testing problems. In the following, $X$ denotes a random sample from the dataset with the smaller number of samples.\\
    
    Recall the proof strategy described after the statement of Theorem \ref{thm:ising}. Continuing in the same vein, we consider the average risk for the hypothesis testing problem
    \begin{align*}
         H_0: \,\,\,&X^n \sim P^{\otimes n}  \\
         H_1: \,\,\,&X^n \sim Q^{\otimes n} \textrm{ for some $Q \in \mathcal{Q}$} 
    \end{align*} 
    under the uniform prior on $\mathcal{Q}$. Recall that the average risk under this prior must be smaller than $R_{\textrm{cd}}$.\\
    
    By the results of Neyman \& Pearson \cite[Ch. 3]{lehmann2006testing}, we know that the most powerful tests for the above hypothesis test are of the form \begin{equation} L_n \overset{H_1}{\underset{H_0}{\gtrless}} \tau,  \notag \end{equation} where $\tau$ is a positive real number, and \begin{equation}L_n(X^n) := \frac{1}{|\mathcal{Q}|}\sum_{Q\in\mathcal{Q}} \prod_{t=1}^n \frac{\dd Q}{\dd P}(X^{(t)}) \notag \end{equation} is the likelihood ratio of the distribution $P$ versus a  distribution uniformly at random from set $\mathcal{Q}$. We refer to $L_n$ as the likelihood ratio of $P$ versus $\mathcal{Q}$.\\
    
    Note that for discrete distributions such as the Ising model, $\frac{\dd Q}{\dd P} = Q/P,$ and for distributions which admit a density with respect to a Euclidean space, such as the Gaussian, we have $\frac{\dd Q}{\dd P} = f_Q/f_P$, where $f_P$ and $f_Q$ denote the respective densities.\\
    
    Our main technical tool is captured by the following lemma, which allows us to compute lower bounds on $n$ required for $R_{\textrm{cd}}$ to be small by computing upper bounds on the variance of $L_n$ for well-chosen ${P}$ and $\mathcal{Q}$.
    
    \begin{mylem}\label{lem:lowb_main}
        Let $\mathcal{M}$ be a class of MRFs. For every $\delta \in [0,1],$ $P \in \mathcal{M}, \mathcal{Q} \subset \mathcal{M}$ such that $P \not\in \mathcal{Q}$, and  $n = \min(n_1,n_2)$, if $L_n$ is the $n$-sample likelihood ratio of $P$ versus $\mathcal{Q},$ then \begin{equation} R_{\textrm{cd}}(n_1,n_2; \mathcal{M}) \le \delta \implies  \E_{P^{\otimes n}}\left[L_n^2\right] \ge 1 + 4(1-\delta)^2. \notag \end{equation}
        \begin{proof}
            For notational brevity, we let \begin{equation} \widetilde{Q}^{n} := \frac{1}{|\mathcal{Q}|} \sum_{Q \in \mathcal{Q}} Q^{\otimes n}. \notag\end{equation} Let $R^{\mathrm{avg}}_{\mathrm{hyp}}$ be the optimal average risk for the above test with $n$ samples when $Q$ is picked uniformly at random from $\mathcal{Q}$. By the previous discussion, 
           \begin{align*}
                 R_{\mathrm{cd}}(n_1,n_2;\mathcal{M}) \ge R^{\mathrm{avg}}_{\mathrm{hyp}} &= \inf_{\tau \ge 0} \P\left(L_n \ge \tau\mid H_0\right) + \P\left(L_n < \tau\mid H_1\right)\\
                &=  1 - \sup_{\tau \ge 0} \left(\widetilde{Q}^{n}\left(L_n \ge \tau \right) - P^{\otimes n}\left(L_n \ge \tau\right) \right)  \\
                &\ge 1 - d_{\text{TV}}(P^{\otimes n},\widetilde{Q}^{n}) \\
                &= 1 - \frac{1}{2} \int_{x^n \in \mathcal{X}^n}  \left| \frac{\dd \widetilde{Q}^{n}}{\dd P^{\otimes n}}(x^n) -1\right| \,\dd P^{\otimes n}(x^n)\\
                &= 1 - \frac{1}{2}\E_{P^{\otimes n}}\left[|L_n - 1|\right] \\
                &\ge 1 - \frac{1}{2}\sqrt{ \E_{P^{\otimes n}}\left[L_n^2\right] - 1}.
            \end{align*}
            Since $R_{\mathrm{cd}} \le \delta,$ the proof concludes with a simple manipulation of the above inequality.
        \end{proof}
    \end{mylem}
    
    \noindent {\emph{Remark:} The above proof technique can also be applied to obtain structure learning bounds. In particular, suppose we have access to a structure learning algorithm that uniformly over all distributions in $\mathcal{M}$ identifies the correct structure with probability greater than $1 - \delta'$. Then, for $(P, \mathcal{Q})$ such that $G(P) \neq G(Q)$ for all $Q \in \mathcal{Q}$, and under the distribution $\P(H_0) = \P(H_1) = 1/2,$ the above hypothesis test can be solved with probability of error smaller than $\delta'$ by learning the structure of the distributions. However, this must exceed $R_{\textrm{hyp}}^{\textrm{avg}}/2$. Thus, all our change detection bounds, which rely on bounding $R_{\textrm{hyp}}^{\textrm{avg}}$, can be converted to structure learning bounds by doubling the reliability level.}

\subsection{Proof of Theorem \ref{thm:ising_easy}}

Let $P$ be the Ising model with no edges on $p$ nodes, and let $Q_{ij}$ be the Ising model with weight $\lambda \in [\alpha, \beta]$ on the edge $(i,j)$ and no other edges. We let the class $\mathcal{Q} := \{ Q_{ij} : 1 \le i < j \le p\}$, and consider the change detection ensemble $(P,\mathcal{Q})$. Lastly, we set $\eta = \frac{e^{-\lambda}}{e^{\lambda}+ e^{-\lambda}}.$ Note that the n-sample likelihood ratio is \begin{equation} L_n(X^n) = \frac{1}{\binom{p}{2}} \sum_{i < j} \prod_{t= 1}^n 2\left( \eta + (1-2\eta) \mathds{1}\big\{X_i^{(t)} = X_j^{(t)}\big\} \right) . \notag\end{equation}

In order to apply Lemma \ref{lem:lowb_main}, we need to compute the quantity $\E_{P^{\otimes n}}\left[L_n^2\right].$ Let $B_{ij}^t := \mathds{1}\{X_i^{(t)} = X_j^{(t)}\}$. We note that for $i<j$, $u<v$, since $P$ is just $p$ independent Bernoulli distributions together, we have  \begin{align*}
\E_P\left[B_{ij}^t\right] &=\,\,\, \frac{1}{4} \quad \textrm{ for every $(i,j)$, and} \\
\E_P\left[B_{ij}^t B_{uv}^t\right] &= \begin{cases} \frac{1}{2} & \textrm{if $(i,j) = (u,v) $} \\ \frac{1}{4} & \textrm{otherwise.}\end{cases}
\end{align*}

Now, \begin{align}\label{eqn:ising_easy_proof}
\E_{P^{\otimes n}}\left[L_n^2\right]   &= \frac{1}{\binom{p}{2}^2} \E_{P^{\otimes n}} \left[ \left(  \sum_{i<j}\prod_{t= 1}^n 2\left( \eta + (1-2\eta) B_{ij}^t\right)\right)^2  \right] \notag\\
&=\frac{1}{\binom{p}{2}^2} \sum_{i<j} \sum_{u<v} \E_{P^{\otimes n}} \bigg[\bigg(\prod_{t= 1}^n 4\Big( \eta^2 + \eta(1-\eta)\left( B_{ij}^t + B_{uv}^t\right) + (1-2\eta)^2 B_{ij}^tB_{uv}^t\Big)\bigg)  \bigg] \notag \\
&\overset{\mathrm{(i)}}= \frac{1}{\binom{p}{2}^2} \sum_{i<j} \sum_{u<v} \Big( 4\eta^2 +  4\eta(1-\eta)\mathbb{E}_P\left[ B_{ij}^1 + B_{uv}^1\right]+ 4(1-2\eta)^2 \mathbb{E}_P\left[B_{ij}^1B_{uv}^1\right]  \Big)^n\notag \\
&\overset{\mathrm{(ii)}}= \frac{1}{\binom{p}{2}^2} \sum_{i<j} \sum_{u<v} \left( 1 + (1-2\eta)^2 \mathds{1}\{(i,j) = (u,v)\} \right)^n \notag \\
&= 1 + \frac{(1 + (1-2\eta)^2 )^n - 1}{\binom{p}{2}}
\end{align}
where the equality $\mathrm{(i)}$ is since each of the $t$ samples are independent and have the distribution $P$, and $\mathrm{(ii)}$ comes from feeding in the moments computed before. Plugging equation $(\ref{eqn:ising_easy_proof})$ into the condition imposed by Lemma \ref{lem:lowb_main} we get that if the risk is smaller than $\delta,$ then we must have \begin{equation}n \ge \frac{1 + 4(1-\delta)^2 \binom{p}{2}}{\log (1 + (1-2\eta)^2)}. \notag \end{equation} We conclude by noting that $1- 2\eta = \tanh (\lambda),$ and that we may set $\lambda = \pm \alpha$. \hfill $\Box$ 

\subsection{Proof of Theorem \ref{thm:Gaussian}}

This result essentially follows the same arguments as in the proof of Theorem \ref{thm:ising_easy}. We identify the distributions of the Gaussian MRFs with their precision matrices. Let $P$ be the Gaussian MRF on $p$ nodes with no edges and unit variance, $Q_{ij}$ be the unit variance Gaussian MRF on $p$ nodes with the single edge $(i,j)$ with edge weight $\lambda$ s.t. $|\lambda| \ge \gamma$, and $\mathcal{Q} := \{Q_{ij}: 1\le i < j \le p\}.$ For convenience, we let $\Delta_{ij}$ be the matrix with the $(i,j)$ and $(j,i)$ entries equal to $\lambda$, and all other entries $0$. Note that the precision matrices are $P = I_p$ and $Q_{ij} = I_p + \Delta_{ij}$. 

Again, consider the likelihood ratio. We have \begin{align}  L_n(X^n) = \frac{1}{\binom{p}{2}} \sum_{i<j} \prod_{t = 1}^n \frac{f_Q}{f_P}(X^{(t)}) = \frac{\left(\det(I + \Delta_{12}\right)^{n/2}}{\binom{p}{2}} \sum_{i<j} \prod_{t = 1}^n e^{ -\frac{\left(X^{(t)}\right)^\tpose \Delta_{ij} X^{(t)}}{2}}. \notag \end{align}

We first require a few computations. For convenience, let $C(\lambda) = \begin{pmatrix} 1 & \lambda \\ \lambda &1 \end{pmatrix}$ and let $i<j$ and $u<v$ be naturals in $[1:p]$\begin{itemize}
    \item[(a)] $\det(I_p + \Delta_{ij}) = 1 - \lambda^2$ - simply observe that the matrix is similar to the block diagonal matrix $\diag(C(\lambda) , I_{p-2}),$ and thus the determinant is the product of $\det(C(\lambda)) = 1 - \lambda^2$ and $\det(I_{p-2}) = 1$.
    \item[(b)] $\det(I_p + 2\Delta_{ij}) = 1 - 4\lambda^2,$ since this is similar to $\diag(C(2\lambda), I_{p-2}).$
    \item[(c)] If $|\{i,j\} \cap \{u,v\}| = 0$, then $\det(I_p + \Delta_{ij} + \Delta_{uv}) = (1- \lambda^2)^2,$ since the two $\Delta$s each contribute separate blocks of $C(\lambda)$.
    \item[(d)] If $|\{i,j\} \cap \{u,v\}| = 1$, then $\det(I_p + \Delta_{ij} + \Delta_{uv}) = 1 - 2\lambda^2$, since this matrix is similar to $\diag(D, I_{p-3})$ where $D = \begin{pmatrix} 1& \lambda & \lambda \\ \lambda & 1 & 0 \\ \lambda & 0 & 1 \end{pmatrix}.$
    
\end{itemize}
We are now in a position to bound $\mathbb{E}[L_n^2]$. The first few steps are parallel to the Ising model case, and are omitted.
\begin{align}
    \E_{P^{\otimes n}}\left[L_n^2\right] &= \frac{\left(\det(I + \Delta_{12})\right)^{n}}{\binom{p}{2}^2} \sum_{i<j}\sum_{u<v} \left(\mathbb{E}_P\left[ \exp{  - \frac{X^\tpose \left(\Delta_{ij} + \Delta_{uv}\right) X }{2} }\right] \right)^n \notag\\
                                           &\overset{\mathrm{(i)}}= \frac{\left(\det(I + \Delta_{12})\right)^{n}}{\binom{p}{2}^2} \sum_{i<j}\sum_{u<v} \left( \det(I_p + \Delta_{ij} + \Delta_{uv} )\right)^{-n/2} \notag \\
                                            &= \frac{1}{\binom{p}{2}^2} \Bigg(\binom{p}{2} \binom{p-2}{2}  + 2\binom{p}{2}\binom{p-2}{1} \left(\frac{1 - \lambda^2}{\sqrt{1 - 2\lambda^2}}\right)^n  + \binom{p}{2} \left( \frac{1- \lambda^2}{\sqrt{1-4\lambda^2}}\right)^n \Bigg) \notag
\end{align} 
where $\mathrm{(i)}$ is due to the Gaussian integral, and the final equality is by simple counting. Note first that $1 - 2\lambda^2 > 1 - 4\lambda^2$. Since we are looking for an upper bound, we will set $a= \frac{1 - \lambda^2}{\sqrt{1 - 4\lambda^2}}$. We thus have  \begin{align}\label{eqn:gaussian_L}
    \E_{P^{\otimes n}}\left[L_n^2\right] \le& \frac{1}{\binom{p}{2}} \left( (2p-3) a^n + \binom{p-2}{2}\right).
\end{align}

Plugging (\ref{eqn:gaussian_L}) into the condition from Lemma \ref{lem:lowb_main}, if the risk is smaller than $\delta$, it must be true that \begin{equation} a^n \ge 1 + 4(1-\delta)^2 \frac{p(p-1)}{2(2p-3)} > 1 + (1-\delta)^2 p. \notag \end{equation}

Taking logarithms, we directly have that for every $|\lambda| \le \gamma,$ \begin{equation} R_{\textrm{cd}}(n_1,n_2;\mathcal{N}_{p,d}(\gamma)) \le \delta \implies n > \frac{\log 1 + (1-\delta)^2 p}{\log \frac{1- \lambda^2}{\sqrt{1-4\lambda^2}}}. \notag\end{equation} The stated result follows on noting that  $\ln (1 - \lambda^2) - \frac{1}{2} \ln (1- 4 \lambda^2) \le 2\lambda^2$ for $|\lambda| \le 0.39$\footnote{This inequality can be shown for $|\lambda| \le \frac{\sqrt{3 + \sqrt{7}}}{2} \approx 0.297$ by a standard positivity of derivative argument, and can be readily shown for $\lambda \le 0.39$ by any numerical equation solver.}, and since if $\gamma \le 0.39, $ we may set $\lambda  = \pm \gamma$ above. Note that the quadratic denominator in $\gamma$ is retained as long as $\gamma$ is bounded away from $1/2$, although with a graceful weakening of the constants invovlved. \hfill $\Box$


\subsection{Proof of Theorem \ref{thm:ising_1}}

To begin with, we will prove a likelihood ratio upper bound for a specific change detection ensemble on the graph class $\mathcal{G}_{d+1,d}$. We will next show a technique that allows us to obtain a closely related bound on the graph class $\mathcal{G}_{p,d}$, and follow this by a small section concluding the proof.

\subsubsection{A likelihood ratio bound}

Let $P$ be the Ising model with uniform edge weight $\lambda$ on the graph $K_{d+1}$, the complete graph on $d+1$ nodes. Note that each node in $K_{d+1}$ has degree $d$. Similarly, let $Q$ be the Ising model with uniform edge weight $\lambda$ on the graph $K_d \setminus \{(1,2)\},$ i.e., the same graph as $P$ but with one edge deleted. We let $Z$ be the partition function for $P$, and $Z'$ be the partition function for $Q$.

Arithmetical manipulations show that for $K_{d+1},$ \begin{equation} \sum_{u < v} \lambda x_ux_v = \frac{\lambda}{2}\left(\left( \sum x_u\right)^2 - (d+1)\right). \notag\end{equation} We thus have \begin{equation} P(X = x) = \frac{1}{Z} \exp{ \frac{\lambda}{2} \left(  \left(\sum_{i = 1}^{d+1} x_i \right)^2 - (d+1) \right)} , \notag\end{equation} and that  \begin{equation} Q(X = x) = \frac{Z}{Z'} P(X= x) e^{-\lambda x_1 x_2}. \notag\end{equation}

We first compute a bound on the $n$ sample likelihood ratio for $P$ versus $Q$, denoted $L_n$. Observe that \begin{align} L_n(X^n) &= \left(\frac{Z}{Z'}\right)^n \prod_{t = 1}^n e^{-\lambda X_1^{(t)}X_2^{(t)}}, \notag\\
\E_{P^{\otimes n}}\left[ L_n^2 \right] &=   \left(\frac{Z}{Z'}\right)^{2n}  \left(\E_P\left[ e^{-2\lambda X_1^{(1)}X_2^{(1)}}\right]\right)^n, \notag
\end{align} and we thus need a bound on $\mathbb{E}_P\left[ e^{-2X_1X_2}\right]$. This is the subject of the following lemma, the proof of which is relegated to the appendix.

\begin{mylem}\label{lem:ising_likelihood} If $d \ge 4$ and $\lambda(d-3) \ge \log d,$ then \begin{equation}	\left(\frac{Z}{Z'}\right)^2 \E_P\left[ e^{-2X_1X_2}\right] \le 1 + 8\left(e^{4\lambda} + d\right)e^{-2\lambda d} \ . \end{equation}
\end{mylem}

\subsubsection{Lifting results on \(d+1\) nodes to \(p\) nodes} \label{sec:K_d_extend}


Let $P$ be a MRF on $d+1$ variables such that $G(P) = K_{d+1}$, and $Q$ be the MRF obtained by eliminating the dependence in $P$ between $X_1$ and $X_2$, i.e, clipping one edge from $G(P)$ with no change in the rest of the graph. Suppose that we have an estimate for the likelihood ratio variance of the form \begin{equation} \E_{P^{\otimes n}}\left[L_n^2 \right] \le h(n,d) \notag \end{equation} where $h$ is some positive function.\\

Let $r := \left\lfloor \frac{p}{d+1} \right\rfloor, $ and consider the graph $G$ on $p$ nodes consisting of $p - r (d+1)$ singletons, and $r$ non-trivial completely connected components, each of which has exactly $d +1$ nodes. Observe that $G = \left(\bigoplus_{i = 1}^r K_d^i\right) \oplus N$, where for $i\in[1:r],$  $K_{d+1}^i$ is the complete graph on the nodes labelled $[1 + (i-1)(d+1): i(d+1)],$ and $N$ is the trivial graph on nodes labelled $[p - r(d+1) + 1, p].$\\

Let $\overline{P}$ be the Markov random field that associates a copy of $P$ with each non-trivial connected component of $G$. In effect, we are sitting with $r$ independent copies of $P$. Further, for $\mu \in [1:r],$ let $\overline{Q}_\mu$ be the MRF that associates a copy of $P$ to each non-trivial component of $G$ except the $\mu^{\textrm{th}},$ with which it associates a copy of $Q$, and let $\overline{\mathcal{Q}}$ be the set of these $\overline{Q}_\mu$s. Suppose $X$ is the random vector associated with the whole graph. We refer to the $d+1$-dimensional random vector that consists of the $X_i$s corresponding to $i$s in the $\mu$th component of $G$ as $X(\mu)$. Finally, we set $\overline{L}_n$ to the likelihood ratio of $\overline{P}$ versus $\overline{\mathcal{Q}}$. Note that \begin{equation} \overline{L}_n = \frac{1}{r}\sum_{\mu = 1}^r L_n(X^n(\mu)), \notag \end{equation} and thus
\begin{align*}
    \E_{\overline{P}^{\otimes n}}\left[L_n^2\right] &= \frac{1}{\rho^2} \sum_{\mu = 1}^\rho \sum_{\nu = 1}^\rho \E_{\overline{P}^{\otimes n}}\left[L_n(X^n(\mu)) L_n(X^n(\nu))\right] \\
    &= \frac{1}{\rho^2} \sum_{\mu = 1}^\rho \sum_{\nu = 1}^\rho \left(\mathbb{E}_{P^{\otimes n}}\left[X^n(\mu)\right]\mathbb{E}_{P^{\otimes n}}\left[X^n(\nu)\right]\right)\mathbbm{1}_{\{\nu \neq \mu\}} \\ &\qquad \qquad \qquad \qquad \quad  + \left( \mathbb{E}_{P^\otimes n}\left[\left( X^n(\mu) \right)^2\right]\right) \mathbbm{1}_{\{\nu = \mu\}}.
\end{align*}
 where the final equality is because non-identical components are independent, and identically distributed according to $P$ in the null hypothesis. Now observe that $\mathbb{E}_{P^{\otimes n}} \left[ L_n(X(\mu))\right] = 1$. Consequently, \begin{align}
    \E_{\overline{P}^{\otimes n}}\left[L_n^2\right] &= 1 - \frac{1}{r} + \frac{1}{r} \mathbb{E}_{P^{\otimes n}} \left[ L_n^2 \right] \notag \\
                                                    &\le 1 + \frac{h(n,d) - 1}{\lfloor p / (d+1) \rfloor}  ,\notag
\end{align} which, by Lemma \ref{lem:lowb_main}, leads to the change detection lower bound \begin{equation}\label{eqn:trick}
R_{\mathrm{cd}}(n_1,n_2;\mathcal{M}') \le \delta \Rightarrow h(n,d) \ge 1 + 4(1-\delta)^2 \left\lfloor \frac{p}{d+1} \right\rfloor, \end{equation} where $\mathcal{M}'$ is a class into which $\overline{P}$ and $\overline{Q}_\mu$ fit.\\


\subsubsection{Finishing Up}

We preserve the notation from the previous two subsections. Let $V:= 1 + 8\left(e^{4\lambda} + d\right)e^{-2\lambda d}$. Recall that by Lemma \ref{lem:ising_likelihood}, \begin{equation} \mathbb{E}_{P^{\otimes n}} \left[ L_n^2\right] \le V^n. \notag \end{equation} 
We form an ensemble on $p$ nodes as in \S\ref{sec:K_d_extend},  and use equation $(\ref{eqn:trick})$ with $h(n,d) = V^n$ and with $\mathcal{M}' = \mathcal{I}_{p,d}(\alpha, \beta)$, to get that for every $\lambda \in [\alpha, \beta] \cap [\frac{\log d}{d-3} , \infty),$ \begin{align*}
\delta &\ge R_{\textrm{cd}}(n_1,n_2;\mathcal{I}_{p,d}(\alpha, \beta)) \\ \implies V^n &\ge 1 + 4(1-\delta)^2 \left\lfloor \frac{p}{d+1} \right\rfloor \\
\iff n &\ge \frac{\log 1 + 4(1-\delta)^2 \left\lfloor \frac{p}{d+1}\right\rfloor}{\log 1 + 8(e^{4\lambda} + d^2) e^{-2\lambda d}}.
\end{align*}

The proof is completed by recognising that for positive $x,$ $(\log 1 + x)^{-1} \ge 1/x,$ and noting that if $\beta(d-3) \ge \log d,$ then we may set $\lambda = \beta$ above. \hfill $\Box$

\section{Discussion}
Our results, both independently and when coupled with the existence of algorithms with sample complexities depending only on the sparsity of changes and not on the maximum degrees of the underlying network structures, raise a few points of interest to both theorists and practitioners. We lay these out below as avenues for future work to explore. 

\begin{enumerate}
    \item As \cite{liu2017learning,liu2017} state, it is believed that the change detection schemes in \cite{FazBan, liu2017, zhang2012learning,zhao2014direct} are agnostic to the underlying models (modulo technical conditions). However, given our results, the technical conditions required for these results to hold cannot include the entirelty of $\mathcal{G}_{p,d}$, since any reliable change detector requires $\Omega\left(\frac{d^2}{(\log d)^2} \log p\right)$. It thus becomes important from the perspective of design and utilisation of change detection algorithms to precisely characterise the technical conditions made in these papers, and which graph classes they in-/preclude.
    \item  Following the initial work by Santhanam \& Wainwright \cite{SanWai}, lower bounds on the structure learning sample complexities for Ising models in a number of graph classes were determined by Tandon et. al. in \cite{TandomShanmugamRavikumarDimakis} via clever refinements and extensions of the original methods. Most of these happen to be exponential in the graph class parameters allowed, and grow at least super-linearly with the same when the model parameters are cherry picked to minimise the lower bounds. On the other hand, it has long been known that the structure learning of Ising models on trees is relatively easy. One then is led to wonder if a similar separation of `easiness' driven by the graphical class of the models extends to the change detection problem, and if so, whether it differs from the parallel separation in structure learning. Concretely, we ask the questions -  \begin{enumerate}
        \item For which graph classes is change detection as hard as structure learning? Again, due to the existence of schemes such as those in \cite{liu2017learning, FazBan}, change detection cannot be as data hungry as structure learning for every graph class. It would be interesting to identify the structural features that make the problem easy.
        \item For which graph classes is change detection easy, in the sense of sample complexities depending only on  sparsity of changes? Is this class larger than the ones determined by the previous work? Further, do larger changes always increase sample costs for detection? Given the practical needs for change detection algorithms, these questions are of significant importance.
    \end{enumerate}
\end{enumerate}

\section*{Acknowledgement}

The authors would like to express their gratitude to Or Ordentlich for many valuable discussions.

\appendix
\section{Proof of Lemma \ref{lem:ising_likelihood}}\label{appx:Ising_bound}

The proof essentially relies on a simple counting and bounding trick. We first demonstrate this trick to obtain an upper bound on $Z$. This is not directly used in the proof, but is the simplest way to demonstrate the result. This is followed by a quick bound on $Z/Z'$, and then a bound on $\E_P\left[ e^{-2X_1X_2}\right].$

\begin{enumerate}
        \item[a)] Recall that $Z$ is the partition function for a complete $d+1$-clique with uniform edge weight $\lambda$. We assume, for simplicity, that $d$ is even, although the bound below also holds for odd $d$.
        \begin{align*}
            Z &= \sum_{\{\pm 1\}^{d+1}} e^{\frac{\lambda}{2} \left( \left(\sum x_i\right)^2 - (d+1) \right)}  \\
            &= \sum_{j = 0}^{d+1} \binom{d+1}{j} e^{ \frac{\lambda}{2} \left( (d+1 - 2j)^2 - (d+1)\right)} \\
           &= 2 \sum_{j = 0}^{d/2} \binom{d+1}{j} e^{ \frac{\lambda}{2} \left( (d+1 - 2j)^2 - (d+1)\right) }.
        \end{align*}

        Here, $j$ is the number of nodes that are $-1$. Let $T_j$ be the $j^{th}$ term in the second eqution above. Observe that as $j$ increases, the $\binom{d+1}{j}$ factors in $T_j$ increase polynomially, but the $e^{\frac{\lambda}{2}(d+1-2j)^2}$ factors decrease exponentially. Thus, for large $\lambda, d$, we expect that an exponentially large fraction of the sum is contributed by the $0^{th}$ term. \\
        
        {In order to show this, let, for $y \in [0,d/2],$ \[\tau(y) :=  \log (d+1 - y) - \log(1+y) - 2\lambda (d- 2y).\] Note that at integer values of $y$, $ \tau(y) = \log \left({T_{y+1}}/{T_y}\right) ,$ and that $\tau(y) < 0$ means that $T_{y+1} < T_y$. By simple computation, it is easy to see that the derivative $\tau'(j)$ is a concave function with the maximum over the considered domain at $d/2.$ In particular, this maxima is positive for $\lambda d \ge 1$. If we now enforce $\tau(0) <0$ by setting appropriate bounds on $\lambda, d$, we find that $\tau$ either decreases and then increases, or just monotonically increases, and in either case it can have at most one root over $[0, d/2]$. However, $\tau(d/2) = 0,$ which means that for $j \in [0:d/2],$  the successive terms are all non-increasing as long as $T_0 > T_1.$ For reasons of convenience, we also demand $T_1 \ge d T_2.$ All the conditions required above are satisfied when $\lambda(d-2) \ge \log d+1 \ge 1$.} \\
        
        With the above in hand, we have, for $\lambda(d-2) \ge \log d+1$, \begin{equation} 2T_0 \le  Z \le 2(T_0 + T_1 + \frac{d}{2}T_2) \le 2(T_0 + \frac{3}{2}T_1), \notag \end{equation} and we hence have \begin{equation}\label{eqn:Z}
            1 \le \frac{Z}{2\exp{\frac{\lambda}{2} ((d+1)^2 - (d+1))}} \le 1 + 3d e^{-2\lambda d} .
        \end{equation}
        
        \item[b)] Recall that $Z'$ is the partition function for the Ising distribution on $K_{d+1} \setminus (1,2)$ with uniform edge weight $\lambda$. Let $\sigma := \sum_{i = 3}^{d+1} x_i$. We have \begin{align}
            Z' &=  \sum_{\{\pm 1\}^{d+1}} e^{\frac{\lambda}{2}\left( (\sigma + x_1 + x_2)^2 -2x_1x_2 - (d+1) \right)} \notag\\
            &\ge e^{-\lambda} \sum_{\{\pm 1\}^{d+1}} e^{\frac{\lambda}{2}\left( (\sigma + x_1 + x_2)^2 - (d+1) \right)} \notag\\
            &= e^{-\lambda} Z \label{ineq:Z/Z'}
        \end{align}
        
        where the inequality holds since $\lambda \ge 0$ and $x_1x_2 = \pm 1$.
        \item[c)] We now upper bound $V:= \left(\frac{Z}{Z'}\right)^2\E_P\left[ e^{-2X_1X_2}\right].$ 
        
        For convenience, we assume below that $d$ is even, although the upper bound remain valid for odd $d$. Using the technique used in the sum in part a) above, 
        \begin{align}
            V &= \left(\frac{Z}{Z'}\right)^2\frac{2}{Z}\sum_{j=0}^{(d-2)/2} \binom{d-1}{j}  \left(1 + 2e^{-2\lambda(d-2j -2) } + e^{-4\lambda(d-2j -1)}\right)    e^{\frac{\lambda}{2}\left( (d+1-2j)^2 - (d+1) -4\right)} \notag \\ 
            &\overset{\mathrm{(i)}}\le e^{2\lambda}\frac{2e^{-2\lambda}}{Z}\sum_{j=0}^{(d-2)/2} \binom{d-1}{j}  e^{\frac{\lambda}{2}\left((d+1-2j)^2-(d+1)\right)}  \left(1+e^{-2\lambda(d-2j-2)}\right)^2  \notag \\
            &\overset{\mathrm{(ii)}}\le \frac{2}{Z}e^{\frac{\lambda}{2}((d+1)^2-(d+1))}  \left(1+e^{-2\lambda(d-2)}\right)^2 \times \left(1 + 6(d-1) e^{-2\lambda d} \right)  \notag \\
            &\le \left(1+e^{-2\lambda(d-2)}\right)^2 \left(1 + 6d e^{-2\lambda d}\right) \notag \\
            &\overset{\mathrm{(iii)}}{\le} 1 + 8(e^{4\lambda} + d)e^{-2\lambda d},\notag
        \end{align}
        where the inequality
        \begin{enumerate}
            \item[(i)] invokes $(\ref{ineq:Z/Z'})$, and uses $\lambda >0.$
            \item[(ii)] follows the strategy used to upper bound for $Z$ as in (\ref{eqn:Z}), and holds if the $T_j$ are decreasing and $T_1 \ge \frac{d}{2} T_2,$ which is true if $ \lambda(d-3) \ge \log d$.
            \item[(iii)] holds if $d\ge 3,$ and $\lambda(d-2) \ge 1$. \hfill $\Box$
        \end{enumerate}
    \end{enumerate}


\RaggedRight
\printbibliography

\end{document}